# A "Turtle Model" of Food System Transformations: Embracing Citizens´ Diverse Values and Knowledge in Change Processes

**Authors:** *Kaiser, M., Cretella, A., Scherer, C., Lam, M.E.*

## Abstract

We explore the challenges and opportunities of transitioning towards sustainable food systems through the lens of democratic food governance fostering inclusive and systemic transformation. Drawing on concepts of wicked problems and systems thinking, we propose a theory of change represented as a ´turtle model´ that embraces the diversity of citizens´ values and knowledge to highlight multiple avenues of transformation. As quadruple helix innovation and governance hubs, cities can be hotspots for food system transformations. We illustrate this for Dublin, Ireland, where local citizens´ value-based food identities were galvanized to activate ecological awareness and promote sustainable seafood consumption. Within this democratic food governance framework, approaches such as open science, transdisciplinarity, and citizen engagement are fit-for-purpose to engage diverse food actors from government, industry, academia, and civil society in shared dialogue and action to transform food systems.

## 1. Introduction: Governing Urban Food System Transitions

Cities are under pressure to transform their food systems, owing to global developments that are challenging conventional practices to feed urban populations. As of 2018, 55% of the world´s population live in cities, a proportion projected to grow to 68% by 2050 (United

Nations 2018). Food systems are at the forefront of urban political agendas, coloured by diverse political ideologies (Holt Gimenez & Shattuck 2011), owing to urbanization and global population growth trends, coupled with persistent food and environmental challenges that contribute to hunger and malnutrition (FAO 2016) and diet-related chronic diseases, such as obesity, and their concomitant health costs (Witten 2016). Cities thus have become critical sites for addressing food issues, with accelerating academic (Harris *et al.* 2015; Cretella 2016) and practical (Steiniger *et al.* 2020; Westskog *et al.* 2024) interest in urban food systems and policies.

Urban food systems research often promotes a "move away from a global, 'industrialized' food system to a more local (or 'alternative') one" (Harris *et al.* 2015, p.64; see also IPES-Food&ETC Group 2021). Many case studies of urban food systems exist already (Deakin *et al.* 2016). This focus on local / regional food systems – and bottom-up approaches to their sustainability – challenge policy-oriented top-down approaches, such as advocated by the UN Food and Agriculture Organization (FAO 2017) and the EAT-Lancet report (Willett *et al.* 2019). In this context, the European Union has played a central role in shaping urban food system transformations, both through policy frameworks and dedicated research investments. Initiatives like the Farm to Fork Strategy (EC 2020, part of the European Green Deal) also support urban food systems in reducing carbon footprints and promoting local production. EU-funded projects, such as *AfriFOODLinks, CITIES2030, CULTIVATE, Edible Cities Network, FIT4FOOD2030, FOODLINKS, Food Trails, FUSILLI, PUREFOOD,* and *SHARECITY*, reflect EU's investment in urban food systems policy and research. It has led to a corresponding surge in academic publications, technical reports, and policy briefs, facilitating not only practical urban food interventions but also food systems understanding.

Projects such as *FIT4FOOD2030* (https://fit4food2030.eu) and *CITIES2030* (https://cities2030.eu) underscore the potential of cities to catalyse systemic food transformations. *FIT4FOOD2030* deployed the European Commission Food 2030 policy framework and convened "multi-stakeholder Labs" across European cities, regions, and countries to catalyse competence development, awareness-raising, and Research and Innovation policy alignment through systems thinking and Responsible Research and Innovation. Meanwhile, *CITIES2030* employs a multi-actor approach to foster the co-creation of resilient, sustainable, and innovative urban food systems, designing progressive policies, formulating action plans, and orchestrating experiments involving researchers, companies, public authorities, civic societies, and citizens. While their case studies and outputs are informative, these projects lack a generalized theory of change model to tackle food system transformations.

Civil society faces mounting pressures to transition towards sustainable food systems (Hebinck *et al.* 2018; van der Heijden 2024; Cretella 2019). While the need to restructure food systems is a global problem, it is markedly more challenging in cities and urban areas, where the mix of societal pressures and drivers is more varied than in rural environments. Food challenges and transformations are inherently complex, with institutional arrangements, market mechanisms, cultural norms, and citizens' behaviors and values all converging in diverse food decisions. Governance of the knowledge and innovation society connects government, industry, academia, and civil society in a "quadruple helix" (Carayannis and Campbell 2009, 2021; Peris-Ortiz *et al.* 2016), a framework often heralded for managing such transitions. Despite the advantages of multiple helix ecosystem approaches, governance models are often still mired in siloed disciplinary approaches that give insufficient consideration to citizens´ perspectives, particularly their values and identities (but see Lam 2021 for a seafood governance example).

We argue that sustainable transitions in urban food systems must prioritize these often-overlooked normative dimensions. Food, as a basic human necessity and a nexus of social, economic, technological, political, and environmental systems, serves as an ideal testbed for exploring alternative governance models to overcome structural challenges, such as funding discontinuities and conflicting stakeholder interests. We frame urban food challenges as fundamentally problems of value, identity, and knowledge diversity (Lam *et al.* 2019; Lam 2021; Kaiser *et al.* 2021; Kaiser 2024a and 2024b; Scharfbillig *et al.* 2021) and promote post-normal science (Funtowicz & Ravetz 1993, 2025) and systems thinking (Meadows 2008; Page 2021; Rutherford 2020) as pathways to systemic food transformations. From this perspective, we develop two core insights:

(i) Transitions must acknowledge and reconcile the diversity of values, identities, interests, knowledge systems and constraints that food actors bring into the governance process; and

(ii) Value-based visions or "visionary pulls" can be key orienting drivers of transformation.

Next, we adapt a middle-complexity generalized theory of change, the "turtle model" of guiding influencing factors used to guide system innovations in the energy transition (Brand, Urte and Gleich 2017), to map out systemic leverage points for guiding urban food system transformations. We then illustrate its potential to guide orientation processes with an example of seafood innovation in Dublin, Ireland. Finally, we reflect on the potential of collective food visions to guide lasting food system transformation that is socially and environmentally sustainable for diverse communities.

## 2. Urban Food System Transformations as "Wicked Problems": Value, Identity, and Knowledge Diversity

*"Only a food systems approach can identify effective intervention points to accelerate climate action while delivering many co-benefits."*

(Glasgow Food and Climate Declaration, 2020)

The global food system is highly complex (like energy, climate, mobility, public health, *etc*), which implies limited predictability, non-linear trajectories, and uncertain system boundaries. System complexity is exacerbated by several interacting sub-systems, with the whole food system itself connected to other complex systems, *e.g.*, society, economy, environment, politics, and health (Figure 1; reproduced from Kaiser *et al.* 2021). As highlighted by the COVID-19 pandemic, change towards sustainable food systems is necessary (Kaiser *et al.* 2021), and by recognizing these interdependencies, food governance models can shift from linear supply chains to more robust integrated, circular frameworks (Parsons *et al.* 2019).

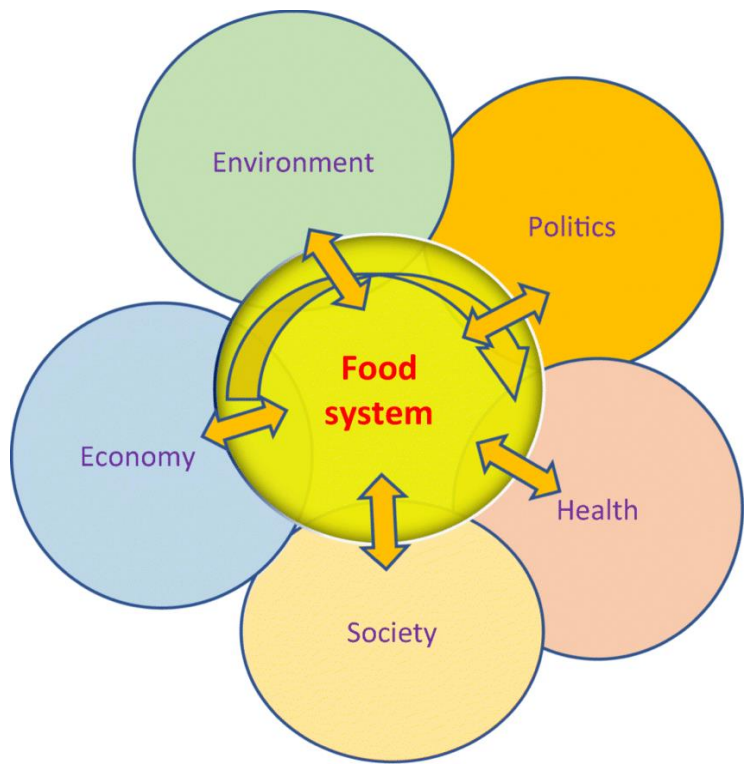

*Figure 1: Circular map of the global food system, with flows connecting the food system with society, economy, environment, politics, and health (reproduced from Kaiser et al. 2021; inspired by, and simplified from Parsons et al. 2019).*

Food system transformations exemplify "wicked problems" (Rittel & Webber 1973; Bentley & Toth 2020), where the problem framing, which already depends on one's values, will bias the proposed solutions. Wicked problems are multi-causal, multi-scalar, and interconnected. They lack a singular problem formulation and true or false solutions: only better or worse answers exist, with no definitive stopping point at which the problem is definitively solved. Thus, wicked problems cannot be tackled by science alone and demand ethical and transdisciplinary approaches that go beyond traditional scientific disciplinary boundaries to reconcile multiple voices and worldviews (*e.g.*, Lam *et al.* 2019). When tackling wicked problems, the scientific and expert community must be vigilant of its own values, paradigms, and knowledge limitations to avoid psychological bias and narrow framings (Kahneman 2011).

To meet the complex food policy challenges that our society faces, we need to think outside the box and break through the disciplinary siloes of normal science. We need to apply many perspectives (Saltelli *et al.* 2020; Page 2021) and do research in a transdisciplinary manner (Lam et al. 2021; Kaiser and Gluckman 2023). Normal science is not good at being concrete nor specific with wicked problems: "Our theories are simply not designed to deal with complex singularities, but rather with simple generalities" (Kaiser 1995, p.195). Therefore, we go "post-normal" to deal with the high complexity, contested values, high system uncertainties, and urgent decision-making needs of real-world problems. Post-normal science offers an approach for navigating these challenges (Funtowicz & Ravetz 1993), emphasizing fitness-for-purpose over absolute truths, and promoting extended peer communities to capture both the inherent and decision-relevant system uncertainties and the socio-political value diversity. It aligns with emerging circular food system models (Jurgilevich *et al.* 2016; Hamam *et al.* 2021), which promote resource interconnectivity and systemic resilience. In the context of food governance, this shift requires explicitly confronting the value conflicts embedded in how problems are framed, whose knowledge counts, and what futures are imagined. Any innovative change will need to be based on these insights (Maye 2018). This orientation paves the way for democratic food governance and the integration of local citizens´ food values, identities, and cultures in systemic food transformation models, to which we now turn.

## 3. Toward Democratic Food Governance: Values and Identities as Orienting Drivers in Systemic Food Transformations

Our suggestion for transforming urban food systems is subtle, but radical: we move from a participatory governance model involving (interest-based) stakeholders to a democratic governance model involving (value-based) citizens. We avoid common assumptions of one-size-fits-all ideologies in nutrition and sustainability science and explicitly embrace diversity in local and regional food systems and diets. We recognize that all societal transition processes start in the minds of people, and it is by addressing their fundamental motivations, that is, their values and beliefs, that people can be motivated to explore more sustainable foodways and to adopt guidelines for healthier diets. By focusing on food values and identities, we highlight orienting processes to potentially robust common future food visions. Within the quadruple helix (Carayannis and Campbell 2009 & 2021) of our democratic food governance framework, we identify citizens as central food actors who can enact common (traditional or future-oriented) food values and identities to re-orient local / regional food systems towards sustainability and ethical acceptance.

In life, our values typically guide what we do: our personal values are our visions of what constitutes a good life, while our social values are our visions of what constitutes a good group or society. Values are constitutive of a person's identity, *i.e.*, who we are and who we want to be are functions of our values. The sociologist Hans Joas (2000, p. 164) claims: "values originate in experiences of self-formation and self-transcendence," where the basic process of the formation of the self is through dialogical experiences. "Identity, not in the sense of stable features, but of a communicative and constructive relationship of the person to himself and to that which does not belong to the self, is the precondition for creative intercourse with the Other and for an ethos of difference," writes Joas (2000, p. 160). Hitlin (2003, p. 123) similarly claims: "values and personal identity are linked at the theoretical level through the concept of authenticity."

The implication for food governance is that food practices and preferences are not only economic or nutritional choices, but also value-laden expressions of one´s identity and belonging. That is, values function much like ideal endpoints for future developments and thus can serve as guides to orient desirable food system transformations. Values and identities are important in policymaking and politics, as highlighted by the EU Joint Research Centre (Scharfbillig *et al.* 2021), and can be mobilized to transform food systems. Recognising this opens alternative, and potentially powerful orienting processes at both the individual and collective levels for systemic food transformations, namely: normative attachments of what people want their food systems to be can guide food transformations beyond instrumental to visionary incentives.

We see values and identities as underutilized potential drivers for food system transformation and hypothesize that a conscious strengthening of local value-based food identities may consolidate transformed dietary choices and long-term practices. Local and regional food identities are strongly influenced by food production methods and practices (Hawkes and Halliday 2017), availability of food resources, and culinary recipes and traditions. While food identities often hark back to childhood experiences, they can also integrate food traditions and innovations specific to a region. For example, stockfish is cultural heritage that connects local food traditions with cultural values and identities in Norway and Italy (Inderhaug 2020), while borscht in Ukraine and bacalao in Portugal contribute to regional / national identities (see: https://slofoodbank.org/en/food-as-culture/). Expanding beyond niche markets to bridge rural and urban areas and local producers and consumers suggests that strengthening shorter supply chains for food (*e.g.*, Lam 2021; Stoll *et al.* 2021), stimulating regional adaptive management of food environments, and providing for a visible regional and value-based food identity that is guiding dietary habits could give regions attractive profiles in tourism and culture, which would benefit local economies.

At the systems level, these value-based food identities can act as a powerful "visionary pull" within food system transformation models, as they embody the futures people want and can provide directionality to change processes. By explicitly engaging with local actors´ values, urban food policies and governance mechanisms can harness food identities as a possible lever for enabling transformation toward more sustainable and resilient food systems. If the UN Sustainable Development Goals (SDGs) are to be met (and not to remain "woefully off the track," as top UN officials stated in 2023 (UN News 2023)), we argue that value-based identities and plural conceptions of the good life must be built into development work to transform urban food cultures toward sustainability.

Specifically, we argue two supportive lemmas:

(i)     All intentional socio-cultural transition processes start in the minds of people.

(ii)    Sustainable futures will be multi-dimensional and must embrace the value and cultural diversity typical of urban environments to catalyse successful transformations.

To engage with such plural, value-driven visions meaningfully, it is necessary to map not only the relevant local actors and their diverse value, identity, and knowledge perspectives, but also the interdependencies and constraints that shape their attitudes, beliefs, and actions. Systems thinking and analysis, to which we now turn, are essential to understand how diverse food drivers, feedbacks, and leverage points interact to either enable or block transformation.

## 4. Thinking in Systems: From Food Chains to Food System Models

Conceptualising food systems in terms of systems thinking allows us to make sense of its inherent complexity, while identifying potential strategic entry points for transformation. Systems thinking simplifies the messiness of highly complex realms of reality to permit classification of the most important drivers and barriers to change of that sector of reality. The process of simplification is a pragmatic necessity to deal intentionally with complexity.

A system consists always of elements, *i.e.*, the domain (*e.g.*, actors: $x$, $y$, and $z$), and relationships between these elements, *i.e.*, the interactions (*e.g.*, $x$ pays $y$; $x$ promotes $y$; $y$

decreases influence of *z*; *etc.*). Some relationships may be uni-directional (*x* → *y*), while others may be bi-directional (*x* ←→ *y*). Furthermore, all systems have a function or, in the case of humans, a purpose. We want to describe the inputs, the outputs, and the processes that characterize the system and that produce the intended functions. These systems can be open (energy and matter can enter), closed (only energy can enter), or isolated (neither energy nor matter can enter). Constructing the system allows a quick overview of the dynamical aspects of the system of interest through an informative visual representation, which we will do for food systems in the next section.

In systems analysis, a system is identified by a boundary separating it from its environment, and its processes are defined in terms of inputs, outputs, and feedback relations. Building a representation or model of a system allows one to analyze and perhaps even steer its development. Any sector of reality will allow several possible representations in system analysis, depending on one's purpose of inquiry. What initiates or blocks processes of change? What determines their envisaged outcomes? Most real systems are complex, composed of numerous interrelated elements: food systems are no exception. By focusing on the major interactions or factors of change, one can identify drivers or leverage points with a high influence on elements and other environmental and social factors that act as barriers to stop or delay processes between elements. A development or theory of change model is then built by identifying major drivers and barriers. This is precisely the aim of the "turtle model" of guiding influencing factors in the next section: to offer a simplified, yet actionable visualisation of how change occurs within food systems, by identifying key elements or actors, the interactions between these actors or the push and pull factors, and the meta factors or barriers constraining the system dynamics.

## 5. The "Turtle Model": A Theory of Change for Urban Food Systems

Our proposed theory of change for food policy development in small- and medium-sized cities adapts the "turtle model" of Brand, Urte and Gleich (2017), who introduced it for guiding orientation processes in system innovations in the energy transition. Our food turtle model (Figure 2) adopts the quadruple helix model of the innovation society to focus on four essential food actors within the socio-economic-political-technical food system: (i) government (constituting the "municipalities") and public agencies (*e.g.*, food banks); 2.

industry (often the engine for innovation and development); 3. academia (the main science and knowledge generation); and 4. civil society, the citizens (constituting "the publics," and including non-governmental organisations or NGOs and civil society organisations or CSOs). Our theory of change for urban food systems focuses on six factors contributing to change: three motivational or push drivers and three orientational or pull drivers, as described below.

Three motivational drivers that can "push" food actors towards changes include: regulative, informational, and technological pushes. Regulative pushes can stem from, for example, international, regional, national and local laws and management regulations governing agricultural and marine resources; environmental, health, and safety standards; and regulations for a green economy. Informational pushes creating negative food perceptions can include: food disgust, scares and scandals; consumer awareness campaigns, often initiated by NGOs; and citizen debates about ethical and sustainable food due to a mismatch between citizens´ consumption habits and dietary and health guidelines or adverse environmental impacts. Technological pushes can originate from technological innovations in the food sector, such as innovations and reforms within the agricultural or seafood production, management, and monitoring systems, novel digitized management and communication tools, and new products based on novel resources, such as insects and algae, as food or ingredients in feed (Cretella 2019).

Three orientational drivers that can "pull" food actors and direct them where to move include: incentive, visionary, and market pulls. Incentive pulls can promote changes via incentives such as government subsidies, funds, prizes, participant opportunities or benefits, and reputational gains related to aspirational goals, such as the UN SDGs. Visionary pulls evoke value-based visions of life quality related to, for example, food identities, health, and happiness, and normative expectations, such as food security, sovereignty, and justice, and environmental and social sustainability. This is the domain of future scenarios, where citizens' and communities' values exert their dominant influence to shape shared visions of a good life. Market pulls include changes in market supply or demand associated with, for example: niche markets; alternative food networks, including shorter supply chains; trusted labels and certification schemes of food quality, ecological sustainability, or fair trade (see, e.g., Lam and Pitcher 2012); food tourism; and circular economy and new green industries.

The whole food system captured by our turtle model is constrained by meta factors or barriers (*e.g.*, climate change, regional and trade wars, power disparities, and shocks) that cannot be easily changed by the food actors, making it difficult to transform the food system. Climate change is shifting local seasonalities (Wardekker *et al.* 2025) and fish migrations patterns (Palacios-Abrantes *et al.* 2025), requiring adapted governance. Inequity across (and within) food systems, accentuated by political tensions among nations influencing food trade and distribution, has led to increasing calls for food justice (Alvarez-Ochoa 2024; Thompson 2016). A heightened demand for greater food system resilience towards possible shocks, such as pandemics or regional wars, is pushing nations towards self-sufficiency (Kaiser *et al.* 2021; Stoll *et al.* 2021). How food systems respond to these external factors depends on intrinsic properties of the system, including the scale and scope of the food networks.

Our food turtle model highlights how different combinations of food drivers can shape distinct transition pathways, depending on local configurations and actor responses. By mapping food actors, push and pull forces, and barriers or constraints together, the food turtle model provides a middle-complexity framework that supports strategic food planning while remaining sensitive to normative dimensions of change. Here, we focus on visionary pull, the capacity of collectively held values and imaginaries to guide transformations towards desired futures. Distinct from incentive and market pulls, visionary pull operates primarily through meaning-making and identity construction. In the next section, we apply our turtle model to an empirical case that illustrates how value-based visions, specifically around food identity, can serve as an effective transformation lever.

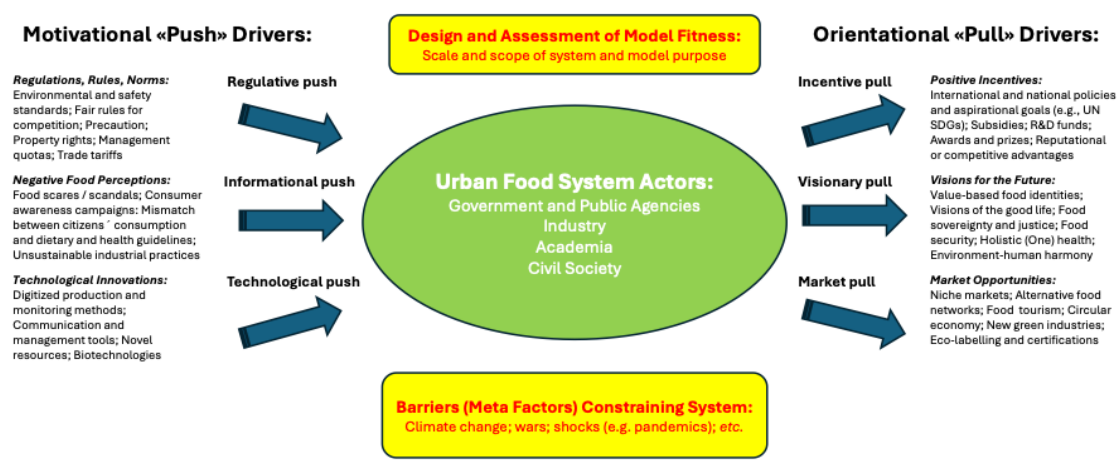

*Figure 2: The Turtle Model for urban food system transformation (adapted for food systems from Brand, Urte and Gleich 2017)*

## 6. Illustrating the Visionary Pull: The Food Smart Dublin Case

The Food Smart Dublin initiative (https://www.tcd.ie/tceh/projects/foodsmartdublin/about/; Scherer & Holm 2020; Cretella, Scherer & Holm 2023) offers a concrete example of how all components of the turtle model can interact to support value-driven food transformation. This initiative in Ireland focused on promoting sustainable seafood consumption by recovering local food heritage, fostering community engagement, and activating ecological awareness. The project was inspired by the SAPEA document Food From the Oceans (SAPEA 2017). It provides an empirical anchor for understanding how push and pull drivers, core actors, and constraints come together in a dynamic urban food system to illustrate how the turtle model can be used in practice for food transformation.

A poster (Figure 3) developed for the Food Smart Dublin project helps summarize its ambition and outreach. It highlights how, despite Ireland's marine territory being ten times its land mass, domestic seafood consumption remains low, particularly of species from lower trophic levels. Irish diets largely focus on predatory fish, such as salmon, cod and tuna, which has adverse ecological and nutritional consequences. The project combined archival research, culinary innovation, and public engagement to motivate more sustainable food choices through six stages: (i) recovering forgotten seafood recipes; (ii) adapting the recipes for modern palates with the help of professional chefs; (iii) involving volunteers recruited by social media to test recipes at home; (iv) collecting feedback through questionnaires; (v) conducting youth workshops with ECO-UNESCO (Ireland's leading environmental education and youth organisation, an NGO); and (vi) publishing an educational cookbook for the public. The initiative is explicitly aligned with the following UN SDGs: Zero Hunger (2); Quality Education (4); Sustainable Cities and Communities (11); Responsible Consumption and Production (12); Climate Action (13); and Life Below Water (14). This project design-structure, combining knowledge production, culinary creativity, and participatory testing, made the societal vision of a sustainable, identity-based seafood culture both tangible and communicable, and hence actionable and realizable.

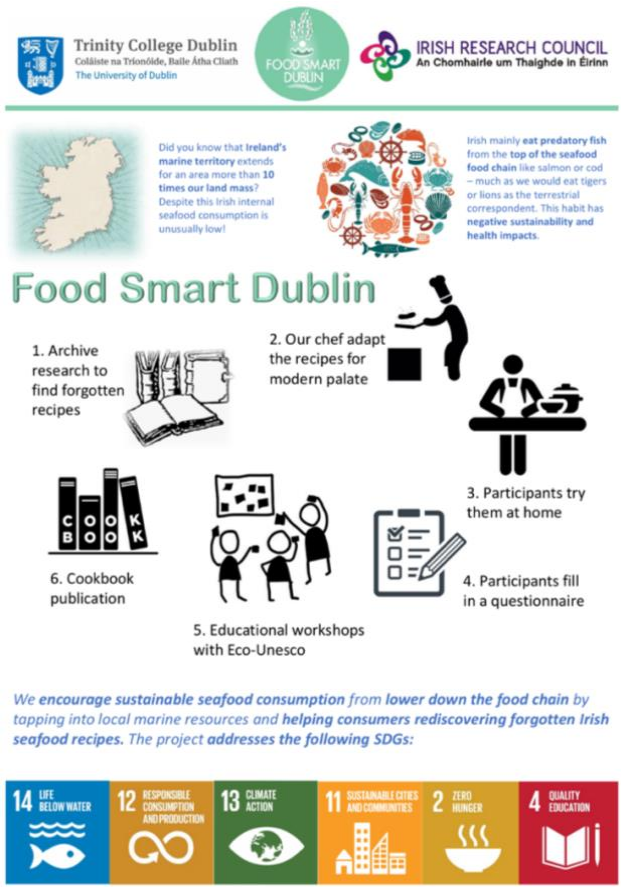

*Figure 3: Poster illustrating the Food Smart Dublin Project (Developed by Agnese Cretella & Cordula Scherer).*

At the centre of the Dublin case were the four main food actor groups identified in the turtle model (see Figure 4): (i) academic institutions (Trinity College Dublin), which led the project design, archival research, and evaluation; (ii) the municipality of Dublin and schools, which provided social-political context and infrastructure; (iii) the food industry, represented by a nationally celebrated sustainable seafood chef and fish monger, Niall Sabongi, and his restaurants, which was central to the re-imagining and dissemination of traditional recipes; and (iv) the public sector, with civil society organisations such as ECO-UNESCO, which helped to engage youth in educational workshops, and citizens, as customers in restaurants or testers of novel dishes. This quadruple constellation of academia, government and schools, industry, and civil society operated in synergy, forming the nucleus of a transdisciplinary and trans-sectoral transformation process. Each food actor group contributed distinct knowledge, values, and constraints, precisely the plurality reflected in the generalized turtle model.

The push drivers in the Food Smart Dublin case were clearly identifiable. Regulative push came through broader alignment with EU food and sustainability frameworks as well as public health agendas. Informational push emerged from growing media and civil society attention to biodiversity loss, unsustainable seafood farming practices and diets, with NGOs and educators helping to translate these concerns into actionable awareness. Technological push was enacted through digital tools and media for public engagement, including updated culinary techniques, online recipe tutorials published on YouTube, and open-access dissemination via the project blog. The recipes and project updates were actively promoted through multiple social media platforms (*e.g.*, Instagram, Twitter, and Facebook), as well as through a dedicated website, newsletter campaigns, and coverage in national newspapers and national radio. These digital tools and communication media not only broadened the project's visibility, but also fostered interaction with a wider public audience, amplifying the informational push.

The major pull factor of Food Smart Dublin was arguably to connect society with its seafood roots. Dublin as a coastal place was conceived as having a sense of belonging to the sea and a strong connection to seafood through coastal sustainable foraging. The project aimed to strengthen this perceived connection to promote Dublin as a coastal city. Therefore, among its pull factors, the visionary pull was dominant. The project mobilised a normative imaginary of Dublin as a sustainable, sea-oriented city—a place where ecological responsibility, food traditions, and social diversity coexist. This vision was realized in recipes, tasting events, storytelling, and youth involvement. Market pull followed, with targeted engagement of younger, environmentally-conscious consumers and culinary innovators. Incentive pull came in the form of reputational value for the restaurants and soft institutional support, including funding and academic prestige for the researchers.

The barriers or meta factors shaping the outer frame of the turtle also were evident. Climate change and marine ecosystem fragility defined the urgency of the project's goals. Power disparities within Ireland's national food policy, especially the dominance of the meat and dairy sectors, posed a structural constraint. Supermarket dominance further reinforced unsustainable consumption patterns by offering a narrow range of high-trophic-level species, such as cod and salmon, while alternative, locally sourced seafood was accessible but apparently less visible or popular to consumers. Finally, the experience of the COVID-19 pandemic underscored the importance of resilient, locally grounded food systems.

Taken together, the Dublin initiative illustrates how the turtle model can structure a systemic, normative, and actor-diverse approach to urban food system transformation. Rather than treating value-based visions as abstract ideals, the project translated them into practice— through participatory design, historical reflection, and culinary experimentation. This shows how "visionary pull" can become not just a conceptual anchor but a practical driver of change.

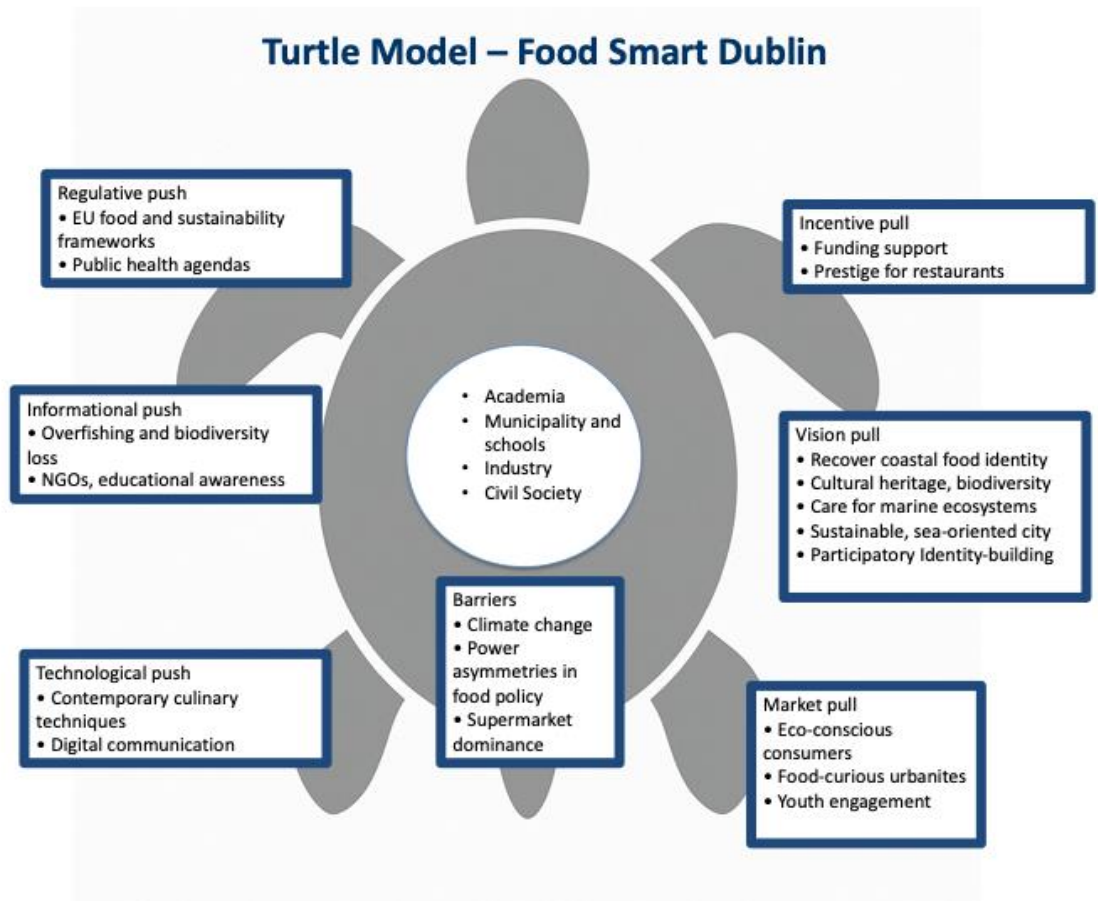

*Figure 4: The Turtle Model applied to the Food Smart Dublin Project*

## 7. Discussion

Transitioning to sustainable urban food systems requires innovative governance that prioritizes systems thinking, citizen participation, and values and knowledge diversity. With their dense concentrations of universities, industries, and civic organizations, cities offer unique opportunities and infrastructure as innovative participatory governance hubs to catalyse systemic sustainability transformations that also can democratize food governance. This can be done by incorporating diverse food actors from government, industry, academia, and civil society, employing the quadruple helix for knowledge and innovation societies, and embracing diverse values, identities, and knowledge, including expert, folk, indigenous, and experiential. Fostering such inclusivity, however, heightens the possibility of tension among competing truth claims and values and also identity politics (Fukuyama 2018). This can be reduced by intentional design following ethical protocols to facilitate rational deliberation and analysis to prioritize resource management and policy tradeoffs (Lam *et al.* 2019; Lam 2021; Kaiser *et al.* 2021).

Within this governance framework, approaches such as open science, transdisciplinarity, and citizen engagement are fit for purpose to transform food systems, as they can foster dialogue that builds shared meaning and trust among diverse food actors to co-create knowledge, align value visions, and innovate future food solutions. Open science promotes participation in, and transparency of the scientific and technological processes and assessments (Dai *et al.* 2018; Potterbusch & Lotrecchiano 2018). Transdisciplinarity (OECD 2020) also focuses on participation and multi-actor network dialogues to develop common problem understandings and cross-cutting methods to address societal objectives (Kaiser and Gluckman 2023, 2025). Citizen engagement, through a co-creation and co-design methodological framework (Jasanoff 2004a, 2004b) in problem formulations, planning, research and development, quality assurance, and implementation, also help to foster trust, epistemic diversity, and value-diversity in outcomes. These approaches, as integrated in post-normal science (Funtowicz & Ravetz 1993), are all oriented towards solving societal and environmental challenges with diverse knowledge inputs and value considerations among all relevant actors, *i.e.*, citizens, academic experts, industry stakeholders, and government representatives (*e.g.*, Lam *et al.* 2019). They aim to improve the quality and resilience of scientific outputs, increase public trust in science, and negate threats to scientific integrity (Kaiser 2014).

By identifying push (motivational) and pull (orientational) drivers in our food turtle model, we present a generalized theory of change that can be used for guiding and analyzing specific food system transformations. We highlight the visionary pull in our Dublin seafood case, which shows how local food actors can be mobilized by their shared food values, identities, and visions to transform food systems to be more sustainable. This is just one example of how the turtle model might be used to guide, that is, orient or "pull" food transformations in a specific local or regional context, but it can also be used to analyze challenges or barriers to transformation. We have limited our turtle model to embody the quadruple helix of innovation, but it can be extended to the quintuple helix (Carayannis and Campbell 2021, Peris-Ortiz *et al.* 2016) by adding the natural environment as core to the innovation process.

Humans evolved shared intentionality (O'Madagain and Tomasello 2019), ingenuity (Gluckman and Hanson 2019), and prospective cognition (Vale et al. 2022) to dominate the planet, but the accelerated pace and severity of impacts of human cultural and technological innovations are overshooting the Earth of its capacity to feed humanity´s growing population (Rees 2023; DeSA 2013). Nowhere is this evident more than in cities, where dense concentrations of urban dwellers increasingly stress and deplete nature´s resources. But the dense and diverse populations of cities make them also incubators of innovation (Harford 2008), so precisely where the problems are the worst is where their solutions are most likely to be found. Communities, in the context of fishing, have been defined by their shared places, histories, and practices (Kraan *et al.* 2025): here, we suggest that urban food communities can be created by harnessing the above human qualities to construct common visions for, and meanings of sustainable and ethical food systems. In transforming urban food systems to reflect not many individuals´ diets, but a community´s values, identity, and goals for the future, we transform the adage "You are what you eat" to "We are what we grow," as a guiding behavioral (re-)orientation of a community towards a shared future vision.

The pragmatics of social innovation towards the SDGs and sustainable and ethical food systems suggest a bottom-up and top-down approach, where our cities can be the testbed for lasting transformative change away from largely unsustainable and unethical food systems.

## *8. Conclusion*

Our paper argues that the salient high complexity of our urban food systems need not hinder our efforts to realize transformative changes in our existing food systems. In our view, it is essential that these efforts are approached from both bottom-up and top-down perspectives, anchored in cities and local communities. Simplifying the complex food system to essential push and pull drivers and barriers can be achieved using systems theory and the turtle model. This provides the basis for participative and normative scenario processes for transformation (Hebinck *et al.* 2018). The attractive feature of the turtle model is its placing of relevant value dimensions in a comprehensive setting which recognizes dominant push and pull influences. Furthermore, it allows the production of multiple future states / scenarios given the typical diversity of values in society. This conceptualization sets the preconditions for a democratic food governance process towards improved regional food systems that citizens can strive for.

In summary, we have introduced the turtle model as a middle-complexity tool well-suited for a democratic food governance approach of transformative change. Its holistic framework and generalizable design elements make it useful as both an analytical tool to interpret, and a pragmatic tool to guide systematic food transformations. This was illustrated by the Dublin Food Smart Initiative, which aimed at sustainable food system change by integrating research activities and engaging citizens within an open-science, transdisciplinary framework. Whether this urban food turtle model approach can be upscaled to food systems within larger segments of society at the national, regional or global scales or be useful within diverse cultural settings, such as in the Global South and Asia, must remain a task for future research.

*  *  *  *  *

## Acknowledgements:

The authors want to acknowledge and thank Poul Holm who led and initiated the Food Smart Dublin project and thus indirectly provided the basis for the early collaboration among two of the authors.

## Declarations:

### Funding sources:

For MK, CS, and AC, no funds, grants, or other support were received for the writing of this paper. MEL was funded by the Research Programme on Marine Resources and the Environment (MARINFORSK) at the Research Council of Norway through the Managing Ethical Norwegian Seascape Activities (MENSA) project (#303663). The cited project Food Smart Dublin in which AC and CS participated was funded by the Irish Research Council.

### Ethics approval and informed consent.

No ethics approval was needed for this research. No participants or informants were consulted. No data were collected since this is a conceptual study.

### Competing interests:

The authors have no relevant financial or non-financial interest to disclose.

### Authors' contributions:

All four authors: Matthias Kaiser (first author), Agnese Cretella, Cordula Scherer, and Mimi E. Lam were involved in the writing of this paper at different stages of its production. Additionally, the first author coordinated the work and acts as corresponding author. All authors share the responsibility for the final version of the paper.

### Clinical trial number:

Not applicable.